\documentclass[10pt,english]{article}
\usepackage[T1]{fontenc}
\usepackage[latin1]{inputenc}
\usepackage{babel}
\usepackage{graphicx}
\usepackage{color}
\usepackage{bm}
\usepackage{longtable}
\usepackage{amsmath} 
\usepackage{amsfonts}
\usepackage{amssymb}
\usepackage{cite}
\usepackage{hyperref}
\usepackage[paperwidth=595.44pt,paperheight=841.68pt,top=72pt,right=72pt,bottom=72pt,left=72pt]{geometry}
\begin{document}
\title{\textbf{High Success Perfect Transmission of 1-Qubit Information Using 
Purposefully Delayed Sharing of Non-Maximally Entangled 2-Qubit Resource and Repeated Generalized Bell-State Measurements
\footnote{Preleminary form of the paper presented at international conference, Quantum Frontiers and Fundamentals, Raman Research Institute, Bengaluru, India. (30$^{th}$ April-4$^{th}$ May, 2018), extended abstract available online,
 http://www.rri.res.in/~qff2018/QFF2018-Abstract.pdf pp.124-126}}}
\author{Shamiya Javed\footnote{javedshamiya@allduniv.ac.in}, Ranjana Prakash\footnote{prakash\_ranjana1974@rediffmail.com} and Hari Prakash\footnote{prakash\_hari123@rediffmail.com}\\Physics Department, University of Allahabad, Prayagraj,  India.}

\date{}
\maketitle
\begin{abstract}
We propose a new scheme in which perfect transmission of 1-qubit information is achieved with high success using purposefully delayed sharing of non-maximally entangled 2-qubit resource and repeated generalized Bell-state measurements (GBSM). Alice possesses initially all qubits and she makes repeated GBSM on the pair of qubits, consisting of (1) the qubit of information state and (2) one of the two entangled resource qubits (taken alternately) until transmission with perfect fidelity is indicated. Alice then sends to Bob, the qubit not used in the last GBSM and also the result of this GBSM and Bob applies a suitable unitary transformation to replicate exactly the information state. We calculate the success probability up to $3^{rd}$ repeated attempt of GBSM and plot it with concurrence of the entangled resource state. We also discuss the maximal average fidelity.
\end{abstract}

\section{Introduction}
Bennett et al \cite{PhysRevLett.70.1895} first proposed a scheme of Standard Quantum Teleportation (SQT) where a sender, say, Alice can send state of a quantum two level system (Qubit) to a distant receiver, say, Bob using a perefectly entangled two-qubit resource \cite{PhysRev.47.777} and a two-bit classical communication channel. Since interaction with environment decoheres the perefectly entangled resource, study of SQT with Non-Maximally Entangled (NME) \cite{cao2006probabilistic,Yan2010,Prakash2012} resource is also important. If the resource is NME, the options available are either to accept imperfect QT with fidelity less than one (maximal average fidelity (MAF)$=(2+C)/3$)\cite{Prakash2012} or to get perfect QT with probabilistic success \cite{PhysRevA.61.034301,AGRAWAL200212,1464-4266-6-8-034}, a process called Probabilistic Quantum Teleportation (PQT). For the latter, Wan Li Li et al \cite{PhysRevA.61.034301} introduced a general form of Bell basis for measurement and showed that for non zero success probability, entanglement of the generalized Bell basis should match with entanglement of the shared resource. Agrawal and Pati \cite{AGRAWAL200212,1464-4266-6-8-034} proposed that using non-maximally entangled generalized Bell basis one can teleport an unknown quantum state via NME resource with unit fidelity and non-unit probability, which is known as PQT. By using basis having same entanglement as that of the NME state success is achieved only in two cases out of four GBSM, with success probability $C^2/2$ (where $C$ is concurrence of the NME resource), which is less than 1/2 for NME resource.
\\\\
In \cite{0253-6102-54-6-12}, the authors considered QT with 2-qubit NME resource and improved the success probability using an ancilla and a controlled rotation operation, in those two attempts in which success is not achieved. For SQT using entangled coherent states \cite{PhysRevA.75.044305,prakash2008effect,Pandey2019,prakash2019controlled}, a scheme of perfect QT with as high success as desired was proposed by Mishra and Prakash \cite{MISHRA2015462}, where entangled coherent states and repeated BSM were used. BSM was a two-stage process, the first indicating whether QT will be  successful or not, and not messing up with the qubit-state, thus permitting repeated BSM's. Improved PQT using 3-qubit resource and suitably prepared ancilla \cite{PhysRevA.61.034301,YAN2003297,Yan2010,0253-6102-54-6-12,2013CoTPh..60..651W,PhysRevA.62.012308,Shi_Biao_2006} and PQT of multi-partite state \cite{SHI2000161,Yong_Jian_2001,0256-307X-19-4-303,PhysRevA.67.014305,0256-307X-23-1-006,hai2006teleportation,article,PATI2007185,XIA2007395,Choudhury2016}  have been proposed and teleportation using mixed entangled state also have been studied \cite{Shi_Biao_2006,li2008probabilistic,prakash2013quantum,adhikari2008teleportation}. The PQT scheme based on cavity QED also has been realized \cite{cao2004probabilistic,cao2006probabilistic}.
\\\\
We propose here a new scheme for transmission of 1-qubit information using NME 2-qubit resource with unit fidelity and high success. This scheme is different from the usual SQT where the 2 entangled qubits of the resource are shared straightway between Alice and Bob and then Alice makes BSM with her two qubits, one have information encrypted on it and the other from the resource state. Our scheme involves purposefully delayed sharing of the two entangled qubits between Alice and Bob and also repeated GBSM's by Alice. Here, Alice gets all qubits initially. She makes repeated GBSM on the pair of qubits, consisting of (1) the qubit having information and (2) one of the two qubit of the entangled resource, chosen alternately, until transmission of information with unit fidelity is indicated. When transmission with perfect fidelity is indicated, Alice sends the qubit not used in the last GBSM and the result of last GBSM (through a 2 c-bit channel) to Bob, who then makes the required unitary transformation on his qubit and gets the exact replica of information. If perfect QT is not indicated, Alice continues with interchange of the qubits of the entangled resource and the repetition of GBSM.

\section{Transmission of 1-qubit information via 2-qubit pure NME resource}
\label{sec:1}

In this section we describe a simple scheme for transmission of 1 qubit information using non maximally entangled 2-qubit resource, with purposefully delayed sharing of the 2 qubits of resource between Alice and Bob and repeated GBSM's by Alice. If Alice has to transmit one qubit quantum information,  $|{I}\rangle_{1}=a|{0}\rangle+b|{1}\rangle$ with $|{a}|^{2}+|{b}|^{2}=1$, to Bob, using NME state, $|{E}\rangle_{23}=\cos{\chi}|{00}\rangle+\sin{\chi}|{11}\rangle$, where $\chi\;\epsilon\;[0,\pi/4]$, with concurrence ${\cal C}=\sin{2{\chi}}$, the combined tripartite state is then given by $|{\psi}\rangle_{123}=(a|{0}\rangle+b|{1}\rangle)_{1}(\cos{\chi}|{00}\rangle+\sin{\chi}|{11}\rangle)_{23}$. Alice possesses initially all the particles and makes GBSM on particles 1 and 2. 
\subsection{Primary Attempt of GBSM}

Alice chooses orthogonal generalized Bell basis as, 
$|{B}^{(0)}\rangle=\cos{\chi}|{00}\rangle+\sin{\chi}|{11}\rangle$, $|{B}^{(1)}\rangle=\sin{\chi}|{00}\rangle-\cos\chi|{11}\rangle$, $|{B}^{(2)}\rangle=\cos{\chi}|{01}\rangle+\sin{\chi}|{10}\rangle$, $|{B}^{(3)}\rangle=\sin{\chi}|{01}\rangle-\cos\chi|{10}\rangle$
and performs GBSM on particle pair (1,2). After measurements the corresponding states of the entangled particle 3 become $N_{1}(a\cos^{2}{\chi}|{0}\rangle+b\sin^{2}{\chi}|{1}\rangle)$, $a|{0}\rangle-b|{1}\rangle$, $a|{1}\rangle+b|{0}\rangle$, $N_{2}(a\sin^{2}{\chi}|{1}\rangle-b\cos^{2}{\chi}|{0}\rangle)$ for GBSM result 0,1,2,3 respectively, where $N_{1}=(|{a}|^{2}\cos^{4}{\chi}+|{b}|^{2}\sin^{4}{\chi})^{-1/2}$ and $N_{2}=(|{a}|^{2}\sin^{4}{\chi}+|{b}|^{2}\cos^{4}{\chi})^{-1/2}$ are normalization constants. Bob's corresponding unitary transformation for these results is $U^{(n)}=\{I,\sigma_z,\sigma_x,\sigma_z\sigma_x\}$.
\begin{table}[h]
\textbf{\caption{GBSM Result in Primary Attempt on particles 1 and 2}}
\centering
\begin{tabular}{l l l c }\\
\hline\\
\textbf{Result} & \textbf{State of particle 3} & \textbf{ Probability} & \textbf{Fidelity}\\[0.5ex]
\hline\\
$|{B^{(0)}}\rangle$ & $N_{1}(a\cos^{2}{\chi}|{0}\rangle+b\sin^{2}{\chi}|{1}\rangle)$ & $P_0=|a|^{2}\cos^{4}{\chi}+|b|^{2}\sin^{4}{\chi}$ & $\neq 1$\\
$|{B^{(1)}}\rangle$ & $a|{0}\rangle-b|{1}\rangle)$ & $P_1=\sin^{2}{\chi}\cos^{2}{\chi}$ & $1$\\
$|{B^{(2)}}\rangle$ & $a|{1}\rangle+b|{0}\rangle)$ & $P_2=\sin^{2}{\chi}\cos^{2}{\chi}$& $1$\\
$|{B^{(3)}}\rangle$ & $N_{2}(a\sin^{2}{\chi}|{1}\rangle-b\cos^{2}{\chi}|{0}\rangle)$ & $P_3=|a|^{2}\sin^{4}{\chi}+|b|^{2}\cos^{4}{\chi}$ & $\neq 1$\\[0.5ex]
\hline
\end{tabular}
\end{table}
\\
it is clear that success is indicated if the GBSM result is $|{B^{(1)}}\rangle$ or $|{B^{(2)}}\rangle$ and failure if the result is $|{B^{(0)}}\rangle$ or $|{B^{(3)}}\rangle$. For indicated success, i.e., GBSM results $|{B^{(1)}}\rangle$ or $|{B^{(2)}}\rangle$, Alice sends the particle not used in the GBSM, i.e., particle 3, to Bob who then applies a suitable unitary transformation to replicate exactly the unknown information state. If Alice's results are $|{B^{(0)}}\rangle$ or $|{B^{(3)}}\rangle$, then she repeats the GBSM but with particle pair (1,3) and not with the earlier used pair (1,2) because a repeated measurement in this case can give nothing new. The calculated success probability in this attempt is, sum of the probabilities for results $|{B^{(1)}}\rangle$ and $|{B^{(2)}}\rangle$ \cite{AGRAWAL200212,1464-4266-6-8-034}, i.e.,   
\begin{equation}
 {\cal P}^{(0)}_{Success}=(P_1+P_2)={\cal C}^{2}/2
 \end{equation}
\subsection{First Repeated Attempt of GBSM}

If Alice finds failure in primary attempt of GBSM, then she can repeat her GBSM on pair (1,3). For failure in the primary attempt, the two cases were primary GBSM results (1) $|{B^{(0)}}\rangle$ and (2) $|{B^{(3)}}\rangle$. We consider these one by one.
\paragraph{Case-(1)}The input state after primary GBSM result $|{B^{(0)}}\rangle$ is $$|{\psi^{(0)}}\rangle_{123}=N_{1}(\cos{\chi}|{00}\rangle+\sin{\chi}|{11}\rangle)_{12}(a\cos^{2}{\chi}|{0}\rangle+b\sin^{2}{\chi}|{1}\rangle)_{3}$$ 
Alice uses orthogonal generalized Bell basis of qubits 1 and 3 as,\\
$|{B}^{(00)}\rangle=(\cos^{3}{\chi}|{00}\rangle+\sin^{3}{\chi}|{11}\rangle)[Y^6(\chi)]^{-1/2}$, $|{B}^{(01)}\rangle=(\sin^{3}{\chi}|{00}\rangle-\cos^{3}{\chi}|{11}\rangle)[Y^6(\chi)]^{-1/2}$, $|{B}^{(02)}\rangle=\cos{\chi}|{01}\rangle+\sin{\chi}|{10}\rangle$, $|{B}^{(03)}\rangle=\sin{\chi}|{01}\rangle-\cos\chi|{10}\rangle$ and perform $1^{st}$ repeated attempt of GBSM on particle pair (1,3).
 
 \paragraph{Case-(2)}
 The input state after primary GBSM result $|{B^{(3)}}\rangle$ is $$|{\psi^{(3)}}\rangle_{123}=N_{2}(\sin{\chi}|{01}\rangle-\cos{\chi}|{10}\rangle)_{12}(a\sin^{2}{\chi}|{1}\rangle-b\cos^{2}{\chi}|{0}\rangle)_{3}$$ 
 Alice uses orthogonal generalized Bell basis as,
$|{B}^{(30)}\rangle=\cos{\chi}|{00}\rangle+\sin{\chi}|{11}\rangle$, $|{B}^{(31)}\rangle=\sin{\chi}|{00}\rangle-\cos\chi|{11}\rangle$, $|{B}^{(32)}\rangle=(\cos^{3}{\chi}|{01}\rangle+\sin^{3}{\chi}|{10}\rangle)[Y^{6}(\chi)]^{-1/2}$, $|{B}^{(33)}\rangle=(\sin^{3}{\chi}|{01}\rangle-\cos^{3}{\chi}|{10}\rangle)[Y^{6}(\chi)]^{-1/2}$
 and perform $1^{st}$ repeated attempt of GBSM on her particle pair (1,3).\\

\begin{table}[t]
\textbf{\caption{GBSM results in first repeated attempt on particles 1 and 3}}

\centering
\begin{tabular}{l l l  c }\\
\hline\\

\textbf{Result} & \textbf{State of particle 2} & \textbf{Probability} & \textbf{Fidelity}\\[0.5ex]
\hline\\
$|{B^{(00)}}\rangle$ & $\sim (a\cos{\chi}|{0}\rangle+b\sin{\chi}|{1}\rangle)$ & $P_{00}=Z^{12}(\chi)[Z^4(\chi)]^{-1}[Y^6(\chi)]^{-1}$  & $\neq 1$\\
$|{B^{(01)}}\rangle$ & $a|{0}\rangle-b|{1}\rangle)$ &$P_{01}=X^{6}({\chi})[Z^{4}({\chi})]^{-1}[Y^{6}({\chi})]^{-1}$ & $1$\\
$|{B^{(02)}}\rangle$ & $a|{1}\rangle+b|{0}\rangle)$ & $P_{02}=X^{4}({\chi})[Z^{4}({\chi})]^{-1}$ & $1$\\
$|{B^{(03)}}\rangle$ & $\sim (a\sin{\chi}|{1}\rangle-b\cos{\chi}|{0}\rangle)$ & $P_{03}=X^{2}({\chi})$ & $\neq 1$\\[0.5ex]
\hline\\
$|{B^{(30)}}\rangle$ & $\sim (a\cos{\chi}|{0}\rangle+b\sin{\chi}|{1}\rangle)$ & $P_{30}=X^{2}({\chi})$ & $\neq 1$\\
$|{B^{(31)}}\rangle$ & $a|{0}\rangle-b|{1}\rangle)$ & $P_{31}=X^{4}({\chi})[Z^{4}({\pi/2-\chi})]^{-1}$ & $1$\\
$|{B^{(32)}}\rangle$ & $a|{1}\rangle+b|{0}\rangle)$ & $P_{32}=X^{6}({\chi})[Z^{4}({\pi/2-\chi})]^{-1}[Y^{6}({\chi})]^{-1}$ & $1$\\
$|{B^{(33)}}\rangle$ & $\sim (a\sin{\chi}|{1}\rangle-b\cos{\chi}|{0}\rangle)$ & $P_{33}=Z^{12}(\pi/2-\chi)[Z^4(\pi/2-\chi)]^{-1}[Y^6(\chi)]^{-1}$ & $\neq 1$\\[0.5ex]
\hline
\end{tabular}\\
\begin{flushleft}
Here symbol $\sim$ denotes that the corresponding state is not normalized and we define the functions of $\chi$, $X^{n}(\chi)=\cos^{n}{\chi}\sin^{n}{\chi}$,  $Y^{n}(\chi)=\cos^{n}{\chi}+\sin^{n}{\chi}$ and  $Z^{n}(\chi)=|a|^{2}\cos^{n}{\chi}+|b|^{2}\sin^{n}{\chi}$.
\end{flushleft}
\end{table}
\begin{flushleft}
Note that in ket for generalized Bell states used in $m^{th}$ repeated GBSM, there are $(m+1)$ digits within brakets. The first refers to result of the original GBSM, the second to result of first repeated GBSM and so on.
\end{flushleft}
Out of eight cases of $1^{st}$ attempt GBSM results Alice can achieve success in four cases. Thus, success probability is increased due to this GBSM attempt $P_0(P_{01}+P_{02})+P_{3}(P_{31}+P_{32})$, the sum of product of probabilities for failure in primary GBSM and the corresponding sums of probability for each success in first repeated GBSM. The success probability at this stage, thus obtained, is \\
\begin{equation}
{\cal P}^{(1)}_{success}={\cal P}^{(0)}_{success}+\frac{{\cal C}^{4}}{8}+\frac{{\cal C}^{6}}{8(4-3{\cal C}^{2})}.
\end{equation}
\subsection{Second Repeated Attempt of GBSM}

The same procedure may be applied when Alice faces failure in $1^{st}$ attempt of repeated GBSM. This time four cases of failure in $1^{st}$ attempt of GBSM are considered.\\
 The possible results in all the above cases are summarized in table-3. Clearly out of 16 cases of  $2^{nd}$ repeated attempt of GBSM, success is achieved in 8 cases. At this stage the final success probability thus obtained is
\begin{table}[h]
\textbf{\caption{GBSM results in second repeated attempt on particle 1 and 2}}
\centering
\begin{tabular}{l l l c }\\
\hline\\
\textbf{Result} & \textbf{State of particle 3} & \textbf{Probability} & \textbf{Fidelity}\\[0.5ex]
\hline\\
$|{B^{(000)}}\rangle$ & $\sim (a\cos^{18}{\chi}|{0}\rangle+b\sin^{18}{\chi}|{1}\rangle)$ & $Z^{36}(\chi)[Z^{12}(\chi)]^{-1}[Y^6(\chi)]^{-1}[Y^{18}(\chi)]^{-1}$ & $\neq 1$\\
$|{B^{(001)}}\rangle$ & $a|{0}\rangle-b|{1}\rangle$ & $X^{18}(\chi)[Z^{12}(\chi)]^{-1}(Y^6(\chi))^{-1}[Y^{18}(\chi)]^{-1}$ & $1$\\
$|{B^{(002)}}\rangle$ & $a|{1}\rangle+b|{0}\rangle$ & $X^{12}(\chi)[Z^{12}(\chi)]^{-1}[Y^6(\chi)]^{-2}$ & $1$\\
$|{B^{(003)}}\rangle$ & $\sim (a\cos^{6}{\chi}|{1}\rangle-b\sin^{6}{\chi}|{0}\rangle)$ & $X^{6}(\chi)[Y^6(\chi)]^{-2}$ & $\neq 1$\\[0.5ex]
\hline\\
$|{B^{(030)}}\rangle$ & $a|{0}\rangle+b|{1}\rangle$ &$X^{6}(\chi)[Z^{4}(\chi)]^{-1}[Y^6(\chi)]^{-1}$ & $1$\\
$|{B^{(031)}}\rangle$ & $\sim (a\cos^{6}{\chi}|{0}\rangle-b\sin^{6}{\chi}|{1}\rangle)$ & $Z^{12}(\chi)[Z^{4}(\chi)]^{-1}[Y^6(\chi)]^{-1}$ & $\neq 1$\\
$|{B^{(032)}}\rangle$ & $\sim (a\cos^{2}{\chi}|{1}\rangle+b\sin^{2}{\chi}|{0}\rangle)$ & $X^{2}(\chi)$ & $\neq 1$\\
$|{B^{(033)}}\rangle$ & $a|{1}\rangle-b|{0}\rangle$ & $X^{4}(\chi)[Z^{4}(\chi)]^{-1}$ & $1$\\[0.5ex]
\hline\\
$|{B^{(300)}}\rangle$ & $a|{0}\rangle+b|{1}\rangle$ & $X^{4}(\chi)[Z^{4}(\pi/2-\chi)]^{-1}$  & $1$\\
$|{B^{(301)}}\rangle$ & $\sim (a\sin^{2}{\chi}|{0}\rangle-b\cos^{2}{\chi}|{1}\rangle)$ & $X^{2}(\chi)$  & $\neq 1$\\
$|{B^{(302)}}\rangle$ & $\sim (a\sin^{6}{\chi}|{1}\rangle+b\cos^{6}{\chi}|{0}\rangle)$ & $Z^{12}(\pi/2-\chi)[Z^{4}(\pi/2-\chi)]^{-1}[Y^6(\chi)]^{-1}$ &  $\neq 1$\\
$|{B^{(303)}}\rangle$ & $a|{1}\rangle-b|{0}\rangle$ & $X^{6}(\pi/2-\chi)[Z^{4}(\pi/2-\chi)]^{-1}[Y^6(\chi)]^{-1}$  & $1$\\[0.5ex]
\hline\\
$|{B^{(330)}}\rangle$ & $\sim (a\sin^{6}{\chi}|{0}\rangle+b\cos^{6}{\chi}|{1}\rangle)$ & $X^{6}(\chi)[Y^6(\chi)]^{-2}$  & $\neq 1$\\
$|{B^{(331)}}\rangle$ & $a|{0}\rangle-b|{1}\rangle$ & $X^{12}(\chi)[Z^{12}(\pi/2-\chi)]^{-1}[Y^6(\chi)]^{-2}$ &  $1$\\
$|{B^{(332)}}\rangle$ & $a|{1}\rangle+b|{0}\rangle$ & $X^{18}(\chi)[Z^{12}(\pi/2-\chi)]^{-1}[Y^6(\chi)]^{-1}[Y^{18}(\chi)]^{-1}$ & $1$\\
$|{B^{(333)}}\rangle$ & $\sim (a\sin^{18}{\chi}|{1}\rangle-b\cos^{18}{\chi}|{0}\rangle)$ & $Z^{36}(\pi/2-\chi)[Z^{12}(\pi/2-\chi)]^{-1}[Y^6(\chi)]^{-1}[Y^{18}(\chi)]^{-1}$  & $\neq 1$\\[0.5ex]
\hline
\end{tabular}
\end{table} 

\begin{equation}
{\cal P}^{(2)}_{success}={\cal P}^{(1)}_{success}+\frac{{\cal C}^{6}}{32}+\frac{{\cal C}^{8}}{32(4-3{\cal C}^{2})}+\frac{{\cal C}^{12}}{32(4-3{\cal C}^{2})^{3}}+\frac{{\cal C}^{18}}{32(4-3{\cal C}^{2})^{3}[4(4-3{\cal C}^{2})^{2}-3{\cal C}^{6}]}
	 \end{equation}
	 To continue the process of achieving as high success as desired, Alice makes $3^{rd}$ repeated attempt of GBSM, 8 cases of failure in $2^{nd}$ repeated GBSM lead to 32 possibilities in $3^{rd}$ repeated GBSM, of which 16 give success. The final success probability up to $3^{rd}$ repeated attempt thus obtained is
	\begin{figure}[h]
\centering
\includegraphics[scale=0.18]{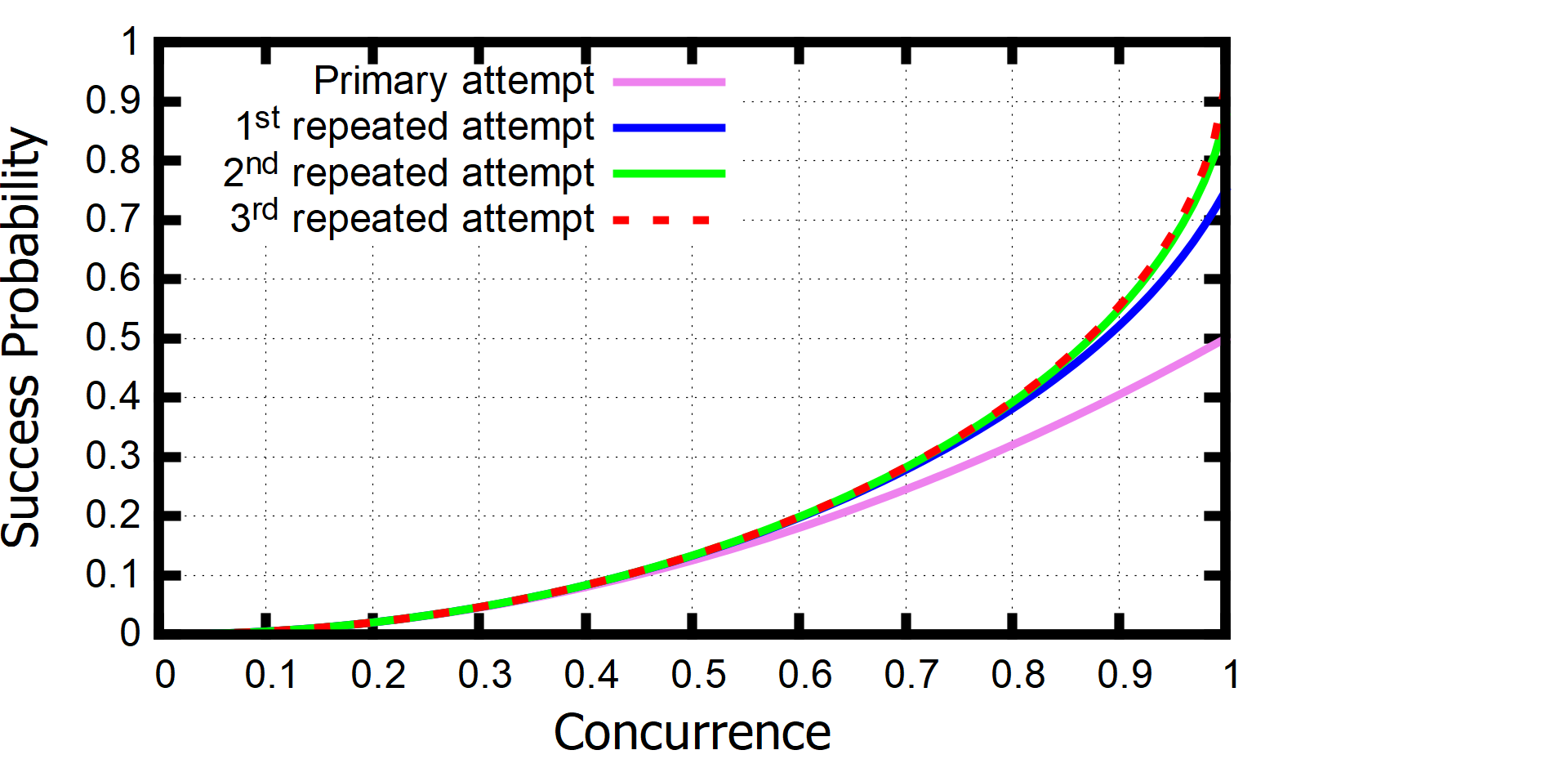}
\caption{\textit{Variation of Success probability with concurrence of the entangled state}}
\end{figure}
\begin{equation}
{\cal P}^{(3)}_{Success} ={\cal P}^{(2)}_{success}+\frac{{\cal C}^{8}}{128}+\frac{{\cal C}^{10}}{128(4-3{\cal C}^{2})}+\frac{{\cal C}^{14}}{128(4-3{\cal C}^{2})^{3}}+\frac{{\cal C}^{18}}{128(4-3{\cal C}^{2})^{5}}+\frac{{\cal C}^{20}}{128(4-3{\cal C}^{2})^{3}[4(4-3{\cal C}^{2})^{2}-3{\cal C}^{6}]}$$$$+\frac{{\cal C}^{24}}{128(4-3{\cal C}^{2})^{5}[4(4-3{\cal C}^{2})^{2}-3{\cal C}^{6}]}+\frac{{\cal C}^{36}}{128(4-3{\cal C}^{2})^{5}[4(4-3{\cal C}^{2})^{2}-3{\cal C}^{6}]^{3}}$$$$+\frac{{\cal C}^{54}}{128(4-3{\cal C}^{2})^{5}[4(4-3{\cal C}^{2})^{2}-3{\cal C}^{6}]^{3}[4(4-3{\cal C}^{2})^{2}((4-3{\cal C}^{2})^{2}-{\cal C}^{6})^{2}-3{\cal C}^{18}]}
	\end{equation}
\\
We plot (figure-1) variation of success probability in each attempt with concurrence of the NME resource state and it is observed that the success increases by repeating GBSM.

	\section{Results and Discussions}
We saw that repeated GBSM can, in principle, lead to SQT with fidelity $F\rightarrow1$ ultimately. But, practically, the number of repetition of GBSM may be limited to, say, $m$ due to some considerations. For this case the choice may be to (1) continue to use the GBSM in the $m^{th}$ repetition or (2) use ME Bell basis in the $m^{th}$ repetition. We find that although (1) may lead to probabilistic success in last attempt (2) leads to greater maximal average fidelity (MAF).

	\begin{figure}[h]
\centering
\includegraphics[scale=0.18]{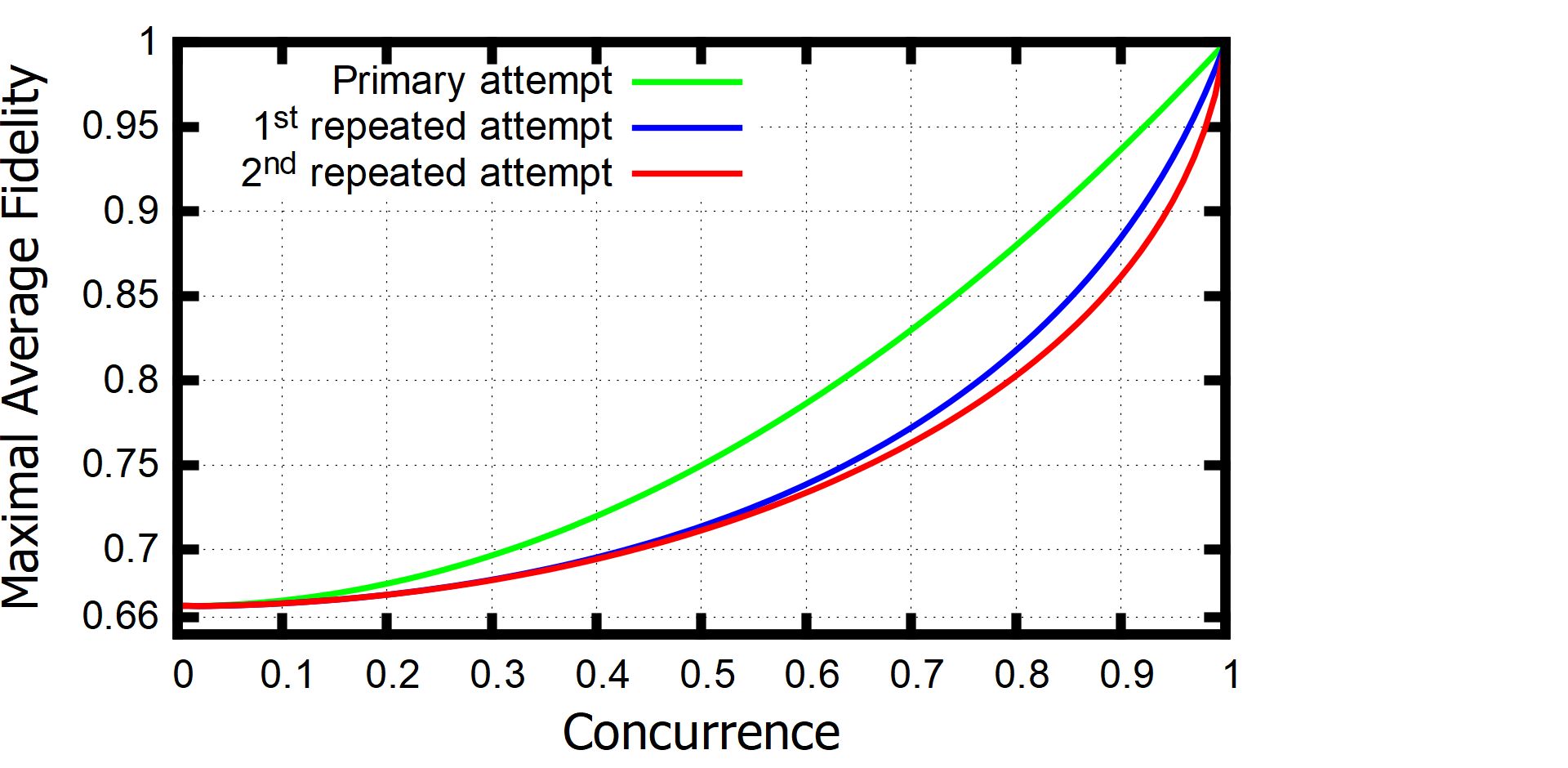} 
\caption{\textit{Variation of maximal average fidelity (MAF) with concurrence of the entangled state with continuation of GBSM}}
\end{figure}	 
\begin{figure}[h]
\centering
\includegraphics[scale=0.18]{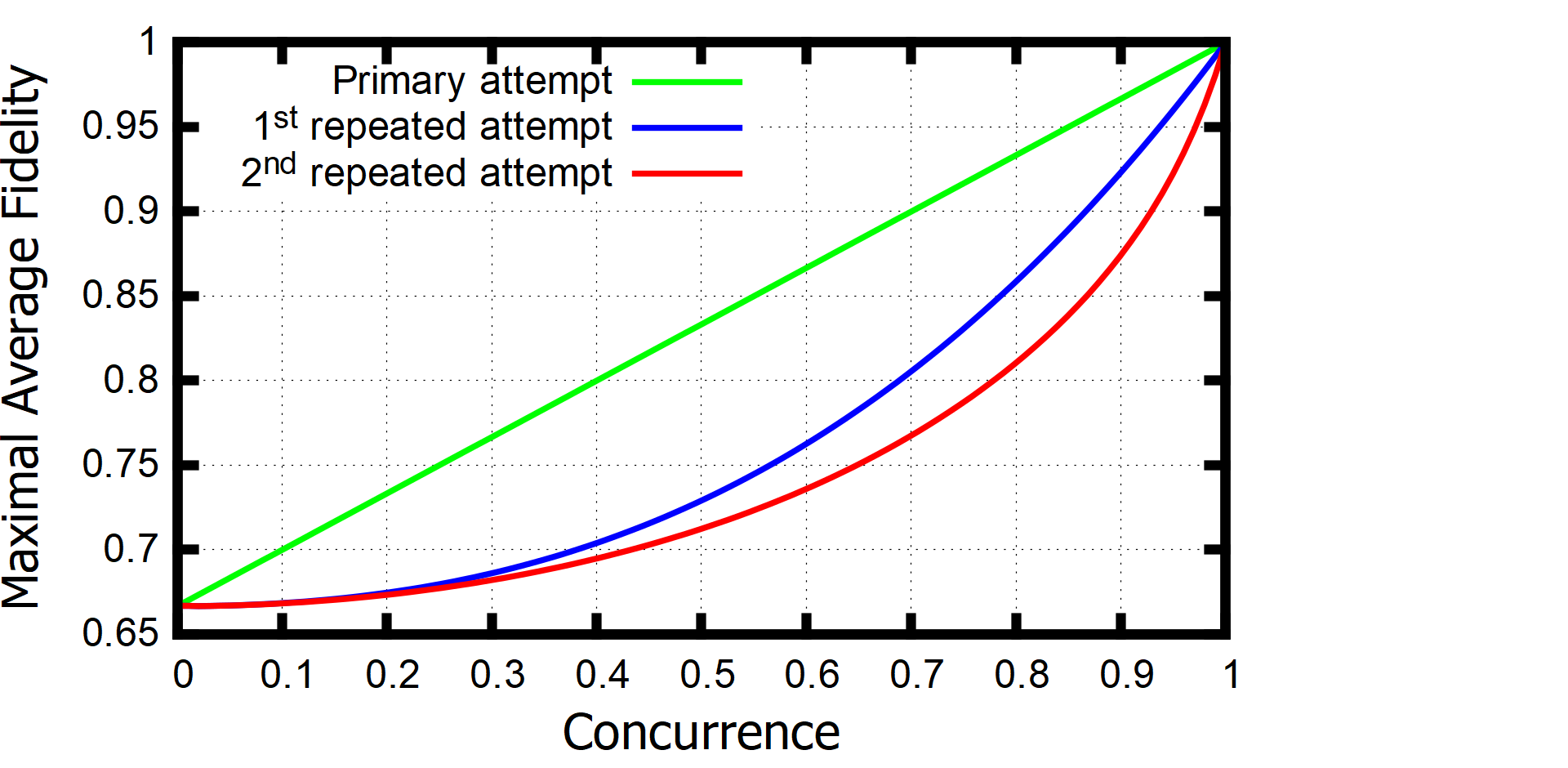} 
\caption{\textit{Variation of maximal average fidelity (MAF) with concurrence of the entangled state with MEBSM at last attempt}}
\end{figure}
 
 We plot the variation of MAF with concurrence for both the above cases as shown in figure-2 and figure-3 respectively. We see that greater MAF is achieved if MEBSM is done. We also note that by increasing $m$ the value of MAF decreases. Hence Alice should be clear in her objective; if she is interested in maximum value of MAF she has to do only one BSM with ME states. However, if she is interested in perfect fidelity with probabilistic success she should do repeated GBSM as shown in this paper.\\
The security in our protocol at every stage is exactly as much as it is in the protocol for standard quantum teleportation (SQT). After BSM, the Bob's state is either $|{I}\rangle$  or one of the three states $\sigma_z|{I}\rangle$, $\sigma_x|{I}\rangle$, $\sigma_x\sigma_z|{I}\rangle$ each with the same probability 1/4. If we consider $|{I}\rangle=a|{0}\rangle+b|{1}\rangle$ and write
$$\rho_{Bob}=\frac{1}{4}[|{I}\rangle\langle{I}|+\sigma_z|{I}\rangle\langle{I}|\sigma_z+\sigma_x|{I}\rangle\langle{I}|\sigma_x+\sigma_z\sigma_x|{I}\rangle\langle{I}|\sigma_x\sigma_z]$$
$$Tr[\rho_{Bob}\rho_{I}]=\frac{1}{2}$$ and if we choose randomly state $|\psi\rangle$  of any qubit, $\rho=|{\psi}\rangle\langle{\psi}|$ and $Tr[\rho\rho_{I}]=\frac{1}{2}$. Thus in quantum teleportation, or in the scheme suggested here, even if Eve manages to get the qubit meant for Bob, she has no means to know the information  ; the only thing she comes to know is that state of qubit with her is either $|{I}\rangle$  or one of the three states $\sigma_z|{I}\rangle$, $\sigma_x|{I}\rangle$, $\sigma_x\sigma_z|{I}\rangle$  and the fidelity of this is $\frac{1}{2}$, which is same as that for any randomly chosen state $|\psi\rangle$.

\section{Conclusions}

 In this paper we show that, in principle, perfect SQT can be done even with an NME resource. In the primary BSM, the Bell state have same concurrence as the resource and the results indicate probabilistic success or failure. For cases of indicated failures, GBSM can be repeated using prescribed NME Bell states. In principle this repetition may be continued till success is achieved.
 If, however, repetition of GBSM is limited to finite number of attempts due to any reason, for probabilistic success with perfect fidelity the GBSM's in the prescribed manner should be done till the end.\\
In this protocol each time the repeated GBSM uses a new more complicated measurement basis, and the price to be paid for improvement in probability for perfect teleportation is complication of measurement basis. Also the security in the suggested protocol is same as it is in the protocol of standard quantum teleportation. So this process of repeated GBSM have some importance to improve success via non-maximally entangled state.  

\section*{Acknowledgement}
 We are grateful to Prof. Naresh Chandra, Prof Phool Singh Yadav and Mr Ravi Kamal Pandey for valuable discussions. One of the author SJ is also thankful to University Grant Commission scheme of Maulana Azad National Fellowship (MANF) for providing financial support.

 \bibliographystyle{unsrt}
 \bibliography{pqt}

\begin{thebibliography}{10}

\bibitem{PhysRevLett.70.1895}
Charles~H. Bennett, Gilles Brassard, Claude Cr\'epeau, Richard Jozsa, Asher
  Peres, and William~K. Wootters.
\newblock Teleporting an unknown quantum state via dual classical and
  einstein-podolsky-rosen channels.
\newblock {\em Phys. Rev. Lett.}, 70:1895--1899, Mar 1993.

\bibitem{PhysRev.47.777}
A.~Einstein, B.~Podolsky, and N.~Rosen.
\newblock Can quantum-mechanical description of physical reality be considered
  complete?
\newblock {\em Phys. Rev.}, 47:777--780, May 1935.

\bibitem{cao2006probabilistic}
Zhuo-Liang Cao, Yan Zhao, and Ming Yang.
\newblock Probabilistic teleportation of unknown atomic states using
  non-maximally entangled states without bell-state measurement.
\newblock {\em Physica A: Statistical Mechanics and its Applications},
  360(1):17--20, 2006.

\bibitem{Yan2010}
FengLi Yan and Tao Yan.
\newblock Probabilistic teleportation via a non-maximally entangled ghz state.
\newblock {\em Chinese Science Bulletin}, 55(10):902--906, Apr 2010.

\bibitem{Prakash2012}
Hari Prakash and Vikram Verma.
\newblock Minimum assured fidelity and minimum average fidelity in quantum
  teleportation of single qubit using non-maximally entangled states.
\newblock {\em Quantum Information Processing}, 11(6):1951--1959, Dec 2012.

\bibitem{PhysRevA.61.034301}
Wan-Li Li, Chuan-Feng Li, and Guang-Can Guo.
\newblock Probabilistic teleportation and entanglement matching.
\newblock {\em Phys. Rev. A}, 61:034301, Feb 2000.

\bibitem{AGRAWAL200212}
Pankaj Agrawal and Arun~K. Pati.
\newblock Probabilistic quantum teleportation.
\newblock {\em Physics Letters A}, 305(1):12 -- 17, 2002.

\bibitem{1464-4266-6-8-034}
A~K Pati and P~Agrawal.
\newblock Probabilistic teleportation and quantum operation.
\newblock {\em Journal of Optics B: Quantum and Semiclassical Optics},
  6(8):S844, 2004.

\bibitem{0253-6102-54-6-12}
Xu~Chun-Jie, Liu Yi-Min, Zhang Wen, and Zhang Zhan-Jun.
\newblock Increasing success probability of a probabilistic quantum
  teleportation.
\newblock {\em Communications in Theoretical Physics}, 54(6):1015, 2010.

\bibitem{PhysRevA.75.044305}
H.~Prakash, N.~Chandra, R.~Prakash, and Shivani.
\newblock Improving the teleportation of entangled coherent states.
\newblock {\em Phys. Rev. A}, 75:044305, Apr 2007.

\bibitem{prakash2008effect}
H~Prakash, N~Chandra, R~Prakash, and Shivani.
\newblock Effect of decoherence on fidelity in teleportation of entangled
  coherent states.
\newblock {\em International Journal of Quantum Information}, 6(05):1077--1092,
  2008.

\bibitem{Pandey2019}
Ravi~Kamal Pandey, Ranjana Prakash, and Hari Prakash.
\newblock Controlled quantum teleportation of superposed coherent state using
  ghz entangled coherent state.
\newblock {\em International Journal of Theoretical Physics}, Jul 2019.

\bibitem{prakash2019controlled}
Ranjana Prakash, Ravi~Kamal Pandey, and Hari Prakash.
\newblock Controlled entanglement diversion using ghz type entangled coherent
  state.
\newblock {\em International Journal of Theoretical Physics}, 58(4):1227--1236,
  2019.

\bibitem{MISHRA2015462}
Manoj~K. Mishra and Hari Prakash.
\newblock Long distance atomic teleportation with as good success as desired.
\newblock {\em Annals of Physics}, 360:462 -- 476, 2015.

\bibitem{YAN2003297}
Fengli Yan and Dong Wang.
\newblock Probabilistic and controlled teleportation of unknown quantum states.
\newblock {\em Physics Letters A}, 316(5):297 -- 303, 2003.

\bibitem{2013CoTPh..60..651W}
J.-H. {Wei}, H.-Y. {Dai}, and M.~{Zhang}.
\newblock {A New Scheme for Probabilistic Teleportation and Its Potential
  Applications}.
\newblock {\em Communications in Theoretical Physics}, 60:651--657, December
  2013.

\bibitem{PhysRevA.62.012308}
Somshubhro Bandyopadhyay.
\newblock Teleportation and secret sharing with pure entangled states.
\newblock {\em Phys. Rev. A}, 62:012308, Jun 2000.

\bibitem{Shi_Biao_2006}
Zheng Shi-Biao.
\newblock Teleportation of quantum states through mixed entangled pairs.
\newblock {\em Chinese Physics Letters}, 23(9):2356--2359, sep 2006.

\bibitem{SHI2000161}
Bao-Sen Shi, Yun-Kun Jiang, and Guang-Can Guo.
\newblock Probabilistic teleportation of two-particle entangled state.
\newblock {\em Physics Letters A}, 268(3):161 -- 164, 2000.

\bibitem{Yong_Jian_2001}
Gu~Yong-Jian, Zheng Yi-Zhuang, and Guo Guang-Can.
\newblock Probabilistic teleportation of an arbitrary two-particle state.
\newblock {\em Chinese Physics Letters}, 18(12):1543--1545, nov 2001.

\bibitem{0256-307X-19-4-303}
Liu Jin-Ming and Guo Guang-Can.
\newblock Quantum teleportation of a three-particle entangled state.
\newblock {\em Chinese Physics Letters}, 19(4):456, 2002.

\bibitem{PhysRevA.67.014305}
Jianxing Fang, Yinsheng Lin, Shiqun Zhu, and Xianfeng Chen.
\newblock Probabilistic teleportation of a three-particle state via three pairs
  of entangled particles.
\newblock {\em Phys. Rev. A}, 67:014305, Jan 2003.

\bibitem{0256-307X-23-1-006}
Yan Feng-Li and Ding He-Wei.
\newblock Probabilistic teleportation of an unknown two-particle state with a
  four-particle pure entangled state and positive operator valued measure.
\newblock {\em Chinese Physics Letters}, 23(1):17, 2006.

\bibitem{hai2006teleportation}
Cao Hai-Jing, Guo Yan-Qing, and Song He-Shan.
\newblock Teleportation of an unknown bipartite state via non-maximally
  entangled two-particle state.
\newblock {\em Chinese Physics}, 15(5):915, 2006.

\bibitem{article}
Cao Min and Zhu Shi-Qun.
\newblock Probabilistic teleportation of multi-particle d-level quantum state.
\newblock {\em Communications in Theoretical Physics}, 43:803, 09 2008.

\bibitem{PATI2007185}
Arun~K. Pati and Pankaj Agrawal.
\newblock Probabilistic teleportation of a qudit.
\newblock {\em Physics Letters A}, 371(3):185 -- 189, 2007.

\bibitem{XIA2007395}
Yan Xia, Jie Song, and He-Shan Song.
\newblock Re-examining generalized teleportation protocol.
\newblock {\em Optics Communications}, 279(2):395 -- 398, 2007.

\bibitem{Choudhury2016}
Binayak~S. Choudhury and Arpan Dhara.
\newblock A probabilistic quantum communication protocol using mixed entangled
  channel.
\newblock {\em Physics of Particles and Nuclei Letters}, 13(3):336--341, May
  2016.

\bibitem{li2008probabilistic}
De-Chao Li and Zhong-Ke Shi.
\newblock Probabilistic teleportation via entanglement.
\newblock {\em International Journal of Theoretical Physics},
  47(10):2645--2654, 2008.

\bibitem{prakash2013quantum}
Hari Prakash and Vikram Verma.
\newblock Quantum teleportation of single qubit mixed information state with
  werner-like state as resource.
\newblock {\em arXiv preprint arXiv:1305.4259}, 2013.

\bibitem{adhikari2008teleportation}
Satyabrata Adhikari, Archan~S Majumdar, Sovik Roy, Biplab Ghosh, and Nilkantha
  Nayak.
\newblock Teleportation via maximally and non-maximally entangled mixed states.
\newblock {\em arXiv preprint arXiv:0812.3772}, 2008.

\bibitem{cao2004probabilistic}
Zhuo-Liang Cao and Ming Yang.
\newblock Probabilistic teleportation of unknown atomic state using w class
  states.
\newblock {\em Physica A: Statistical Mechanics and its Applications},
  337(1-2):132--140, 2004.

\end{thebibliography}

\end{document}